\documentclass{aa}
\usepackage[varg]{txfonts}
\usepackage{graphicx}
\usepackage{amsmath}
\usepackage{natbib}
\usepackage{hyperref}
\usepackage{xcolor}
\usepackage{wasysym}
\usepackage{xspace}
\usepackage{subcaption}
\usepackage{caption}
\usepackage{float}
\usepackage{siunitx}

\bibpunct{(}{)}{;}{a}{}{,} 

\newcommand{\mic}{~$\mu$m\xspace}

\graphicspath{{data/}}

\begin{document}

\title{Radial distribution of the carbonaceous nano-grains in the protoplanetary disk around HD 169142}

\author{Marie Devinat\inst{1}\and
 Émilie Habart\inst{2}\and
  Éric Pantin\inst{3}\and
  Nathalie Ysard\inst{2}\and
  Anthony Jones\inst{2}\and
  Lucas Labadie\inst{4}\and
  Emmanuel Di Folco\inst{5}}
  
\institute{École Normale Supérieure Paris-Saclay, Université Paris-Saclay, 4, avenue des Sciences, 91190, Gif-sur-Yvette, France\and
Université Paris-Saclay, CNRS, Institut d’Astrophysique Spatiale, 91405, Orsay, France\and
IRFU/SAp Service D’Astrophysique, CEA, Orme des Merisiers, Bât 709
91191, Gif-sur-Yvette, France\and
I. Physikalisches Institut, Universität zu Köln, Zülpicherstr. 77, 50937 Cologne, Germany\and
Laboratoire d'Astrophysique de Bordeaux, Université de Bordeaux, CNRS, B16N, allée Geoffroy Saint-Hilaire, 33615, Pessac, France}

\date{2021}

\keywords{<Protoplanetary disks - Infrared: planetary systems>}

\abstract
        {HD 169142 is part of the class of (pre-)transitional protoplanetary disks showing multiple carbon nanodust spectroscopic signatures (aromatic, aliphatic) dominating the infrared spectrum. Precise constraints on the spatial distribution and properties of carbonaceous dust particles are essential to understanding the physics, radiative transfer processes, and chemistry of the disk. The HD 169142 disk is seen almost face-on and thus it offers a unique opportunity to study the dust radial evolution in disks.}
   {We investigate the spatial distribution of the carriers of several dust aromatic emission features of the disk across a broad spatial range  (10-200 au) as well as their properties. }
   {We analysed imaging and spectroscopic observations in the 8–12\mic range from the VLT Imager and Spectrometer for mid-Infrared (VISIR) at the Very Large Telescope (VLT), as well as adaptive optics spectroscopic observations in the 3–4\mic range from the Nasmyth Adaptive Optics System -- Near-Infrared Imager and Spectrograph (NACO) at VLT.
   The data probe the spatial variation of the flux in the 3.3\mic, 8.6\mic, and 11.3\mic aromatic bands. To constrain the radial distribution of carbonaceous nano-grains, the observations were compared to model predictions using The Heterogeneous dust Evolution Model for Interstellar Solids (THEMIS), which is integrated into the POLARIS radiative transfer code by calculating the thermal and stochastic heating of micro- and nanometer-sized dust grains for a given disk structure.} 
   {Our data show predominant nano-particle emission at all radii (accessible with our resolution of about 0.1" or $\sim$ 12~AU at 3\mic and $\sim$ 0.3", 35~AU at 10\mic) in the HD 169142 disk. This unambiguously shows that carbonaceous nano-grains dominate radiatively the infrared spectrum in most of the disk, a finding that has been suggested in previous studies.
   In order to account for both VISIR and NACO emission maps, we show the need for aromatic particles distributed within the disk from the outermost regions to a radius of 20\,AU, corresponding to the outer limit of the inner cavity derived from previous observations. In the inner cavity, these aromatic particles might be present but their abundance would then be significantly decreased.}
   {}

\maketitle

\section{Introduction}

The solid carbon content of protoplanetary disks is of primary importance for improving the understanding of circumstellar disk evolution and planet formation. \citet{Oberg2011} linked the initial carbon content of the region of formation of planetesimals to the composition and chemistry of the atmosphere of the evolved planet. \citet{Eistrup2018} showed that the chemical composition of forming planets depends on the physical conditions and the location of planetesimals in the disk, emphasising the necessity of taking chemical evolution into account
in planetary formation models. Hence, an accurate and complete knowledge of the physical properties of primordial solid carbon reservoirs through the disk is essential for describing the formation and evolution paths and predicting the probability distribution of the properties of evolved planets (e.g. \citet{Oberg2011}). 

If the solid carbon in disks has been extensively studied at large separations, it is only recently that investigations in planet-forming regions, namely, those below $\sim$50~AU, were made possible through high-resolution observations. In tracking down carbonaceous emission lines with the MATISSE interferometer, \citet{Kokoulina2021} were able to probe the distribution of solid carbon below 10~AU in the disk around HD 179218. Recently, \citet{Bouteraon_032019} investigated the spatial distribution of carbon nano-grains down to 0.1" in protoplanetary disks around multiple stars, including HD 100546 and HD 169142. 

\begin{figure}[!ht]
    \centering
    \includegraphics[width=0.5\textwidth]{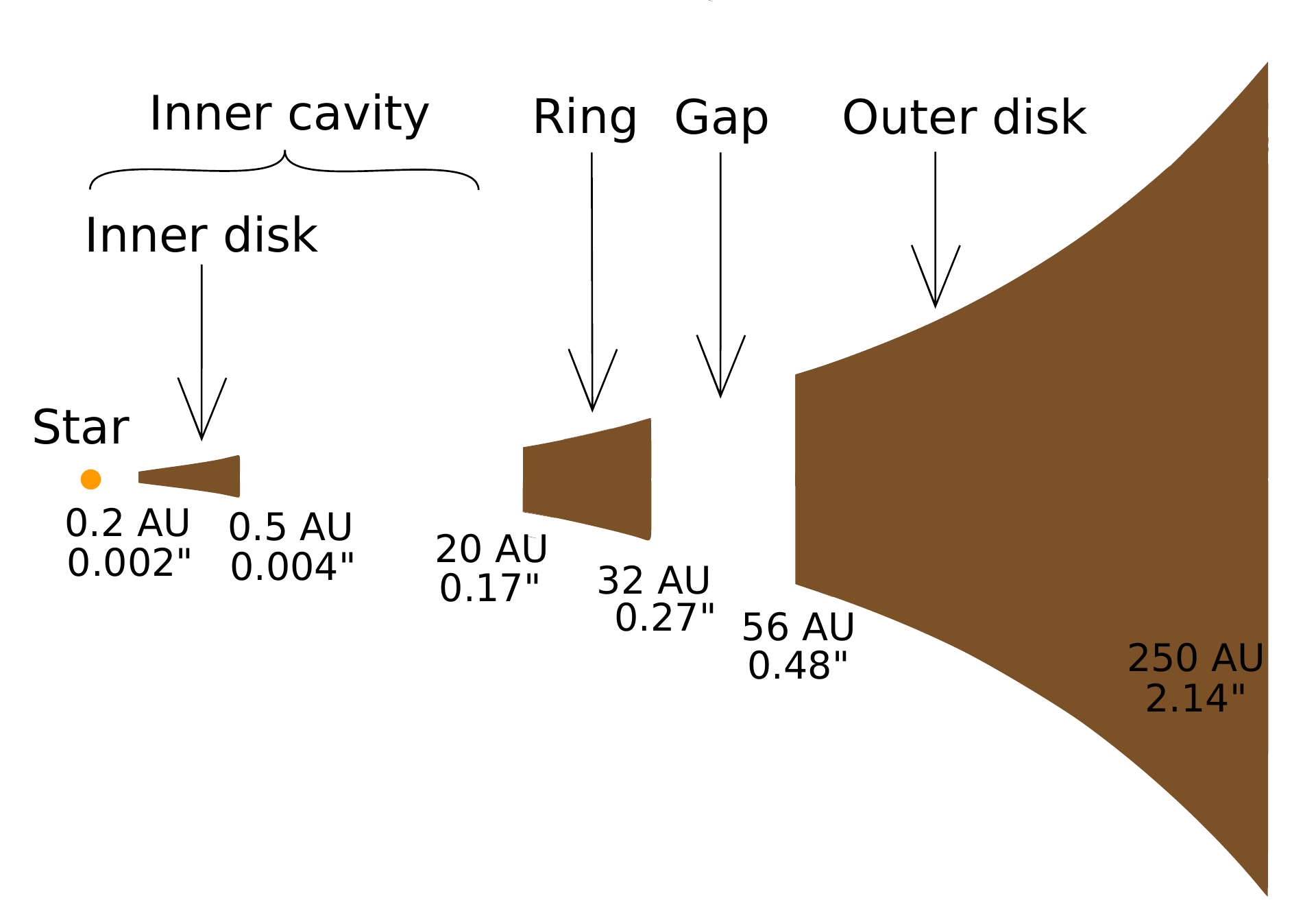}
    \caption{Structure of the disk around HD 169142, taken from \citet{Osorio_2014, Monnier_2017, Ligi2018}.}
    \label{fig:diskstructure}
\end{figure}

In this paper, we study HD 169142, a well-known Herbig Ae star (A5Ve; \citet{Keller2008}), at a distance of 117 $\pm$ 4~pc \citep{Manoj_2006, Grady2007, GaiaCollaboration2016} (see stellar properties in Table \ref{tab:HD 169142_star}). The star is surrounded by an almost face-on (pre-)transitional disk, with an inclination $\sim$~\ang{13} \citep{Panic2008}.
The gas and dust spatial distributions observed with SPHERE and ALMA hint at the presence of multiple (giant) protoplanets shaping the disk structure via dynamical clearing (dust cavity and gap) and dust trapping (double ring dust distribution). 
The protoplanetary system shows an inner disk between 0.2 and 0.5~AU, two annular gaps (from 1 to 20~AU and from 32 to 56~AU), and an outer disk extending to 250~AU \citep{Quanz_2013, Osorio_2014, Ligi2018, Momose2015, Fedele2017, Bertrang2018, Honda2012, Perraut2019}. The radial structure of the disk used in this work is summarized in Fig. \ref{fig:diskstructure}. It assumes the gap radial extension derived by \citet{Ligi2018} from the results of \citet{Quanz_2013}, using the corrected value of 117 pc for the distance to Earth.

HD 169142 is one of the few Herbig Ae/Be stars for which planet candidates have been detected via direct imaging in the near-infrared \citep{Biller_2014, Reggiani_2014, Osorio_2014, Quanz_2013}. One of the point sources detected around HD 169142 \citep{Biller_2014, Reggiani_2014} was shown to be related to an inhomogeneous ring structure in the inner region of the disk at 0.18" by \citet{Ligi2018}. These authors also found that an additional compact structure detected at about 0.1’’ projected separation could possibly be related to an additional ring structure at the given separation. 
Furthermore, the system shows near infrared flux decreasing in the past decades (by 45\%), which may have resulted from modifications of the inner disk structure 
due to accretion or sculpting by planets undergoing formation \citep{Wagner2015}.  

The HD 169142 global spectrum presents very rich and intense infrared carbon nano-dust spectroscopic features (aromatic, aliphatic) between 3 and 13~$\mu$m \citep{Meeus2001, Sloan2005, Acke2010, Bouteraon_032019, Seok2017}. The carriers of these bands can be attributed to carbon nano-grains containing aromatic domains, for instance,
astronomical polycyclic aromatic hydrocarbons (PAH) or amorphous hydrocarbons a-C(:H) that are
undergoing stochastic heating.
HD 169142 is one of the few disks with a very large amount of aromatic species \citep{Woitke_2016} and very high normalised aromatic infrared bands to UV luminosity ratio $L_{AIB}/L_{UV}$ \citep{Acke2004, Maaskant2014}. 
In this study, we present spatially resolved observations of the 3.3, 8.6, and 11.3\mic aromatic bands. 

The disk of HD 169142 has been observed at different wavelengths, using ISO, Spitzer, and ground-based spectroscopy. 
The detection of [OI] \citep{Acke2005}, H$\alpha$ \citep{Dunkin_1997}, and Br$\gamma$ \citep{GarciaLopez2006} emission indicates the presence of gas in the inner parts of the disk. 
The carbonaceous aromatic nano-particles are strongly coupled to the gas in proto-planetary disks and are thus expected to follow the radial distribution of gas in the disk. Given the detected distribution of gas, it is then natural to wonder about the presence of aromatic carbonaceous nano-particles in the inner regions of the disk.

\citet{Habart2006,Bouteraon_032019} detected the 3-4\mic aromatic features in the inner parts of the disk. These features are spatially extended, with a FWHM of 0.3” or $\sim$30\,AU. 
Both the spatial extension of carbonaceous nano-grains in the HD 169142 disk and their presence in the optically thin inner cavity have also been suggested by \citet{Maaskant2014} and \cite{Seok_2016}. Furthermore, \citet{Seok_2016} performed a comprehensive modelling of the SED with a mixture of porous dust and astronomical-PAH and found that three dust populations and relatively small PAH with an ionization fraction of 0.6 can explain the entire SED. 
Backing up those observations, \citet{Klarmann2017} compared the extended continuum flux measured at 1.3\mic with PIONIER \citep{Lazareff2016} and the aromatic infrared bands to UV luminosity ratio in multiple circumstellar objects and found an extended hot emission in the cavity of the HD 169142 disk, which they linked to stochastically heated carbonaceous particles. 
Finally, based on resolved Keck NIRC2 data in the 3.4-4\mic region \citet{Birchall2019} suggested the presence of a ring of emission at $\sim$ 0.06" (7~AU). They conclude that, in contrast with the historical view of pre-transitional discs (e.g. \citet{Espaillat2007}), the ‘gap’ region is radiatively dominated by emission from very small grains and very small, likely carbonaceous, nano-particles. 

In this paper, we use high angular-resolution data obtained with the NACO and VISIR instruments to investigate the spatial distribution of the carriers of the aromatic infrared bands at 3.3, 8.6, and 11.3\mic in the HD 169142 disk. We focus on the distribution of the carriers in the most evolved, innermost regions of the disk (inner cavity and gap, below 60~AU). We model the data using The Heterogeneous dust Evolution Model for Interstellar Solids\footnote{https://www.ias.u-psud.fr/themis/} (THEMIS, \citet{Jones2017}) integrated in the radiative transfer code POLARIS \citep{Reissl2016} to constrain the spatial distribution of carbonaceous nano-particles and to compare it with the distribution of micron- to millimeter-sized dust grains derived in the literature. 
The data are presented in Sect. \ref{sec:observations} and analysed in Sect. \ref{sec:obsresults}. The radiative transfer modelling is described in Sect. \ref{sec:RTmodels}. Section \ref{sec:results} compares it to the data and constrains the radial distribution of aromatic carbon grains. We   discuss and summarize our findings in Sect. \ref{sec:conclusion}.

\begin{table*}[!ht]
        \centering
        \begin{tabular}{ccc}
                \hline
                Parameter&Value&Source\\
                \hline  
                Mass ($\text{M}_{\astrosun}$)& 2; 1.65&[1], [2]; [4], [5], [8]\\
                Age (Myr)&5.4; 3-12; 6; 12&[1]; [4]; [5]; [8] \\
                Distance (pc)&145; \textbf{117}; 151&[1], [2], [4], [5], [6]; [7]; [8]\\
                Radius $R_*$ ($\text{R}_{\astrosun}$)&2.2; \textbf{1.6}&[1], [2]; [5],[8]\\
                Temperature (K)&8\,100; 8\,250; 8\,400; 6\,500; \textbf{7\,800}&[1], [2]; [5]; [6]; [8]\\
                Spectral type&Herbig Ae/Be : A5Ve; A7V; A7Vz&[1], [2], [5], [6]; [4]; [8] \\
                \hline
        \end{tabular}
        \caption{HD 169142 stellar properties. Values are taken from [1] \cite{Manoj_2006}, [2] \cite{Monnier_2017}, [3] \cite{Quanz_2013}, [4] \cite{Biller_2014}, [5] \cite{Seok_2016}, [6] \cite{Dunkin_1997}, [7] \cite{Ligi2018}, [8] \cite{Blondel_2006}. The values chosen for the current study are in bold font.}
        \label{tab:HD 169142_star}
\end{table*}

\section{Observations and data reduction} \label{sec:observations}
We analysed the distribution of the carriers of aromatic infrared bands in the HD 169142, making use of VISIR imaging (Sect. \ref{sec:photometryVISIR}) and spectroscopic observations (Sect. \ref{sec:spectra}), as well as NACO spectroscopic data (Sect. \ref{sec:NACOdata}).

\subsection{VISIR imaging data} \label{sec:photometryVISIR} 
HD 169142 was observed by the ESO mid-infrared instrument VISIR installed on the VLT (Paranal, Chile) in the imaging mode in the PAH1 and PAH2 filters, which probe the aromatic infrared bands at 8.6 and 11.3\mic. In addition, the source was observed in the adjacent continuum (ArIII and SIV filters at 9 and 10.5\mic respectively).
The observations were obtained in 2005 as part of the VISIR GTO program on circumstellar disks, and partly from the VISIR archive (ESO program 099.C-0794(A) on 07/21/2017). All observations were obtained under fair conditions of seeing (less than 1" DIMM monitor) and precipitable water content smaller than 5 mm (Tab. \ref{tab:obslogimg}). Under good seeing conditions, close to diffraction-limited imaging is obtained in the N band (8-13\mic) with an angular resolution of about 0.3" around 10\mic. A standard quadrangular (8" offsets) chopping-nodding technique was used to remove the high thermal background.
All observations except in the ArIII filter used the burst mode, which allows for one image to be stored every $\sim$ 15 ms. A dedicated pipeline was then used to perform lucky imaging \citep{Law2006} and image registration to minimize the effect of tip-tilt smearing. Typical achieved Strehl ratios are of the order of 0.5. Standard calibration stars were also observed in the same modes in order to perform the photometric calibration and determine the instrument PSF (Table \ref{tab:obslogimg}).

\begin{figure}[ht]
    \centering
    \begin{subfigure}{0.25\textwidth}
    \includegraphics[width=1\textwidth]{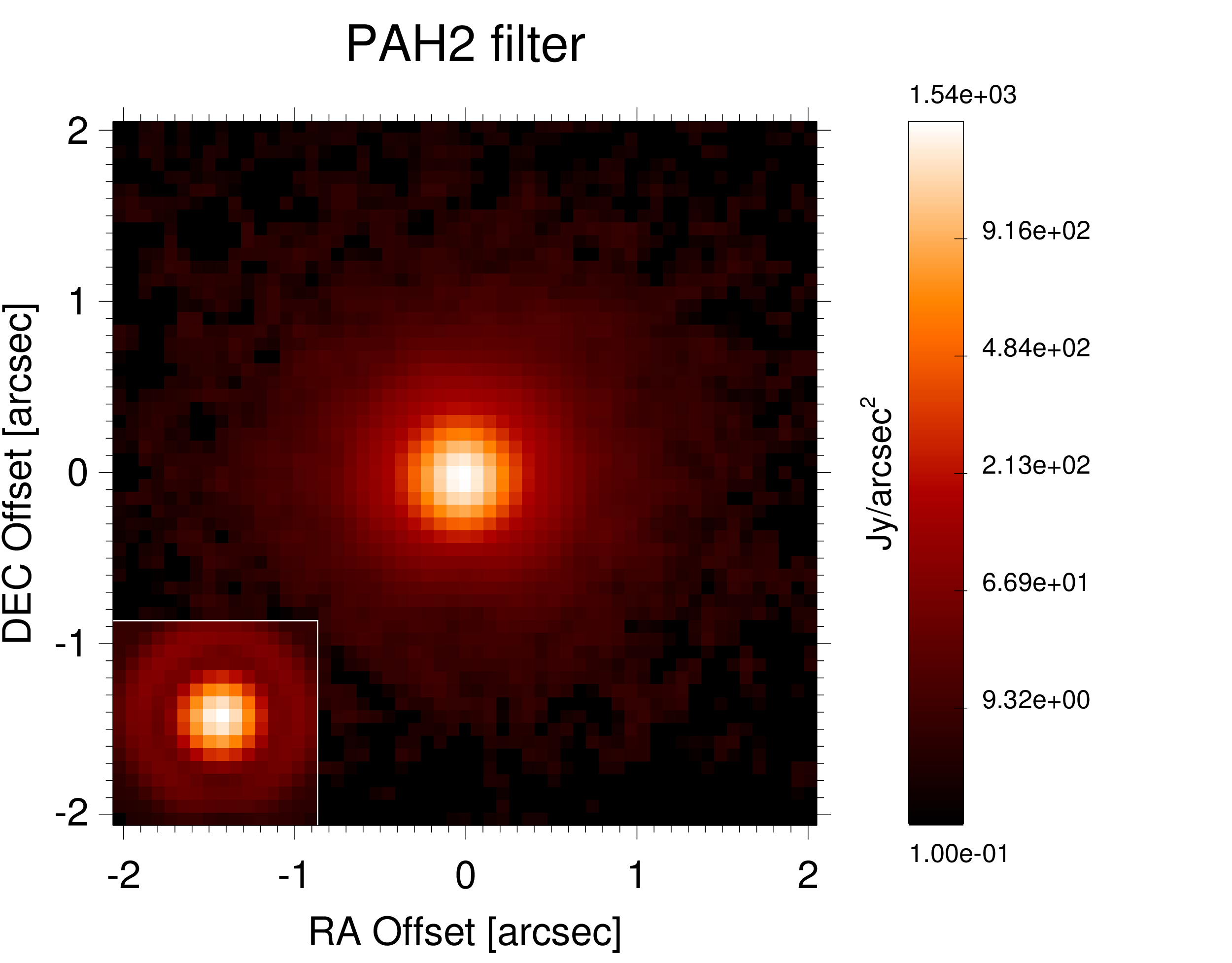}
    \includegraphics[width=1\textwidth]{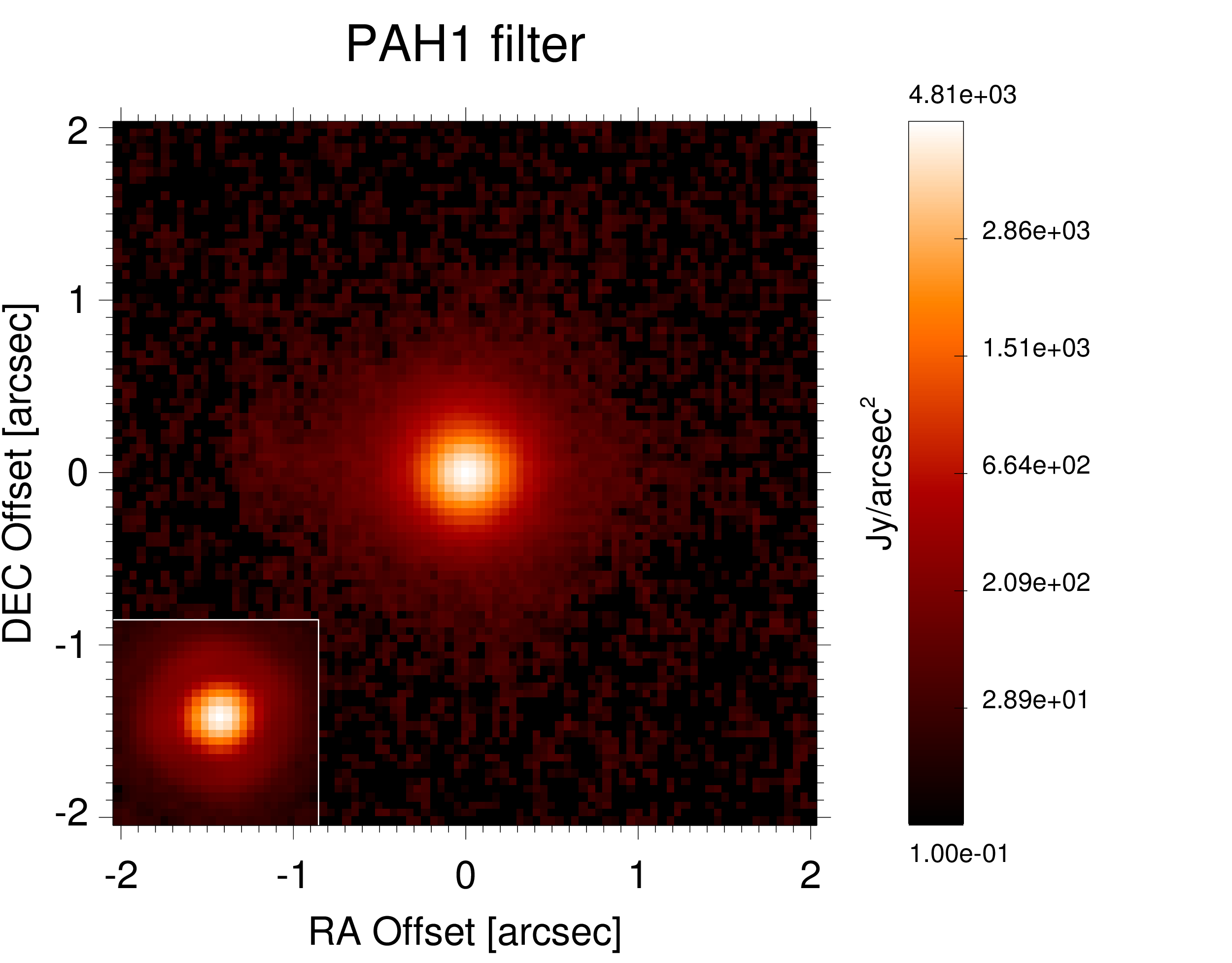}
    \end{subfigure}
    \begin{subfigure}{0.23\textwidth}
        \includegraphics[width=1\textwidth]{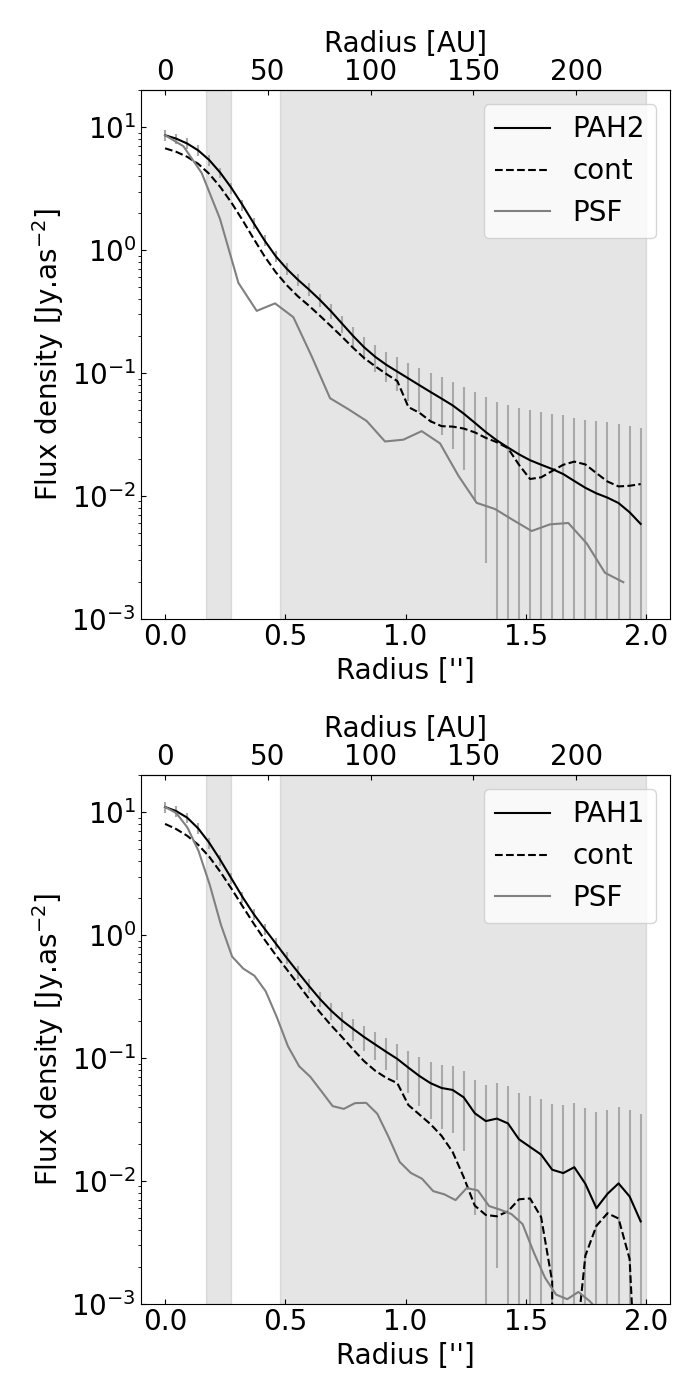}
    \end{subfigure}
    \caption{Total emission map (band + continuum) in the 11.3\mic PAH2 and 8.6\mic PAH1 VISIR filter, compared to the PSF of the observation in the same filters in the inset (\textit{Left panel)}. The PAH1 and PAH2 emission in HD 169142 are spatially more extended than the filter PSF and show no Airy ring, which is also indicative of a spatially resolved emission. Azimuthally averaged emission profiles (centered on the brightest pixel) in the PAH1 and PAH2 VISIR filter, compared to the underlying continuum emission, shown as dashed lines (computed as described in Sect. \ref{sec:continuumderivation}), and the associated PSF profiles, shown as thin lines (Right
panel). The error bars on the profiles, taking into account the absolute calibration error, are shown in grey. The location of micron- and millimeter-sized grain dust, taken from the literature (see Fig. \ref{fig:diskstructure}), are indicated by the grey fill.}
    \label{fig:VISIRphotodata}
\end{figure}

Data taken in PAH2, ArIII, and SIV filters have a pixel scale of 0.075". They are re-binned to the PAH1 pixel size of 0.046" which corresponds to the upgraded VISIR instrument pixel scale after 2015. The emission maps probe the disk from 0" to 3". However, this study is limited to the range [0", 2"] (see Fig. \ref{fig:VISIRphotodata}) as the signal-to-noise ratio (S/N) drops at larger radii.

 \begin{table*}[!ht]
     \centering
     \begin{tabular}{lcccc}
          \hline
          Filter & PAH1 & ArIII & PAH2 & SIV \\
          \hline
          Wavelength [\mic]         & 8.6      & 8.99     & 11.3    & 10.4 \\
          On source obs. time [min] & 10       & 10       & 20       & 20 \\
          Std star                  & HD167121 & HD151680 & HD151680 & HD151680 \\
         \hline
     \end{tabular}
     \caption{Characteristics of the VISIR imaging observations.}
     \label{tab:obslogimg}
 \end{table*}

\begin{table}[!ht]
    \centering
    \caption{Full width at half maximum in arcseconds, measured with the use of gaussian fitting on the VISIR 2D maps in the PAH1 and PAH2 filters (1$^{st}$ column) and the PSF of the observations (second column).}
    \begin{tabular}{cccc}
        \hline
          & Filter FWHM [arcsec] & PSF FWHM [arcsec]\\
        \hline
         PAH1 & 0.41 & 0.26\\
         PAH2 & 0.47 & 0.30\\
        \hline
    \end{tabular}
    \label{tab:fwhms}
\end{table}

\subsection{VISIR spectroscopic data} \label{sec:spectra}
The spectroscopic observations were performed on the night of 18-19June 2005 (ESO program 60.A-9234(A)). The observing conditions were fair at less than 0.8" optical seeing and relatively low humidity.
The low-resolution spectroscopic mode of VISIR was used ($R\simeq220$). The 0.75" slit was  north-south aligned. The entire N band is split into four grating settings. In this paper, we focus only on the ones centred on the aromatic features at 8.6 and 11.3\mic, with the other ones  affected by rather strong telluric transmission errors. The on-source observing time for each setting was about ten minutes, which corresponds to a median sensitivity of about 50 mJy/arcsec$^2$. 
Chopping and nodding were performed parallel to the slit for all observations. The standard star used for photometric calibration and telluric correction was HD 177716 \citep{Cohen1999}.

\begin{figure}[!ht]
    \raggedleft
    \includegraphics[width=0.39\textwidth]{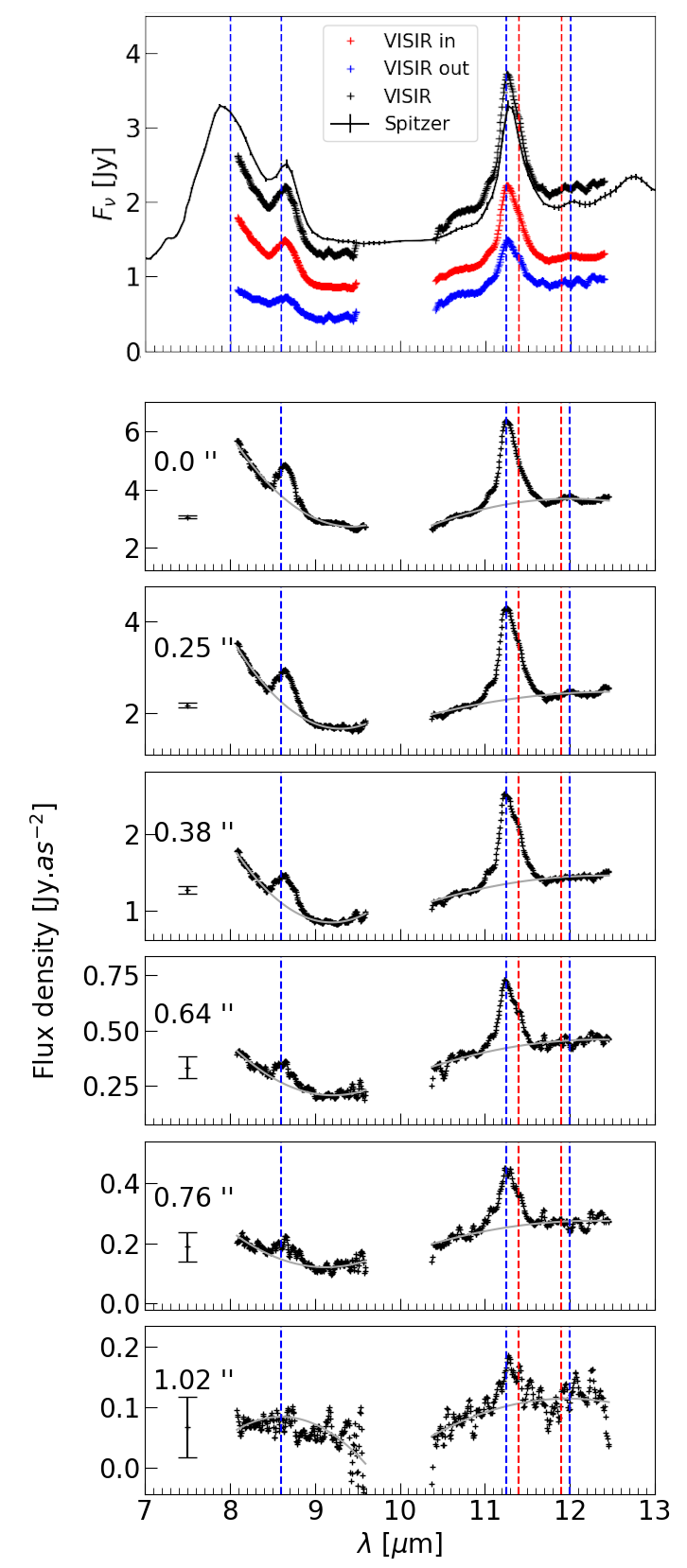}
    \centering
    \caption{HD 169142 integrated spectra measured by Spitzer (crosses), and spectrum summed from all the VISIR resolved spectra (black line, see bottom panels) are shown in the upper panel. The red curve is the sum of VISIR spatially resolved spectra from 0.0" to 0.25", corresponding to the inner 30 AU, and the blue curve is the sum from 0.38" to 1.02", corresponding to the outer disk. Correction for the flux outside of the slit is done by weighting each spectrum by the surface of a ring of corresponding radius, and width, 0.127". Panels 2-7 show the surface brightness is measured on spatially resolved slit spectra at radii of 0", 0.25", \dots The spectra are not continuum subtracted. They probe ring-like regions of a width 0.128" (1 pixel) and mean radii of 0", 0.25" \dots in the disk. The spectral region without data delimits the telluric ozone absorption region. The over-plotted vertical lines show the positions of the aromatic (blue) and crystalline forsterite (red) main features. The typical $\pm$3$\sigma$ error derived from background noise (50~mJy.as$^{-2}$) is shown in every panel. The grey full line shows the continuum interpolation under each band by a 2-degree polynomial.}
    \label{fig:VISspectra}
\end{figure}

VISIR resolved spectra are measured from 0" to 1.5", with a pixel size of 0.127". However, only the data under 1" is used (see Fig. \ref{fig:VISspectra}) because the S/N drops below a value of two at higher radii.
 
\subsection{NACO data} \label{sec:NACOdata}
Observations of the HD 169142 disk were performed between 3.2 and 3.76\mic using a long slit with the adaptive optics system NACO at the VLT. The on-sky projection of the slit is 28" long and 0.086" wide, which corresponds to the diffraction limit in this wavelength range. The  spectral resolution is $R = \lambda / \Delta \lambda \sim 1\,000$. Nine slit positions were taken, one centred on the star and the other slits shifted by half a slit width. Nine slit positions allowed to extract a spectral cube on an area of $2^{\prime\prime}\times0.354^{\prime\prime}$ centred on the star. 
The dataset ESO program ID is 075.C-0624(A). Observations properties and data reduction are summarised in \citet{Bouteraon_032019}.

\begin{figure}[!ht]
    \centering
    \includegraphics[width=0.5\textwidth]{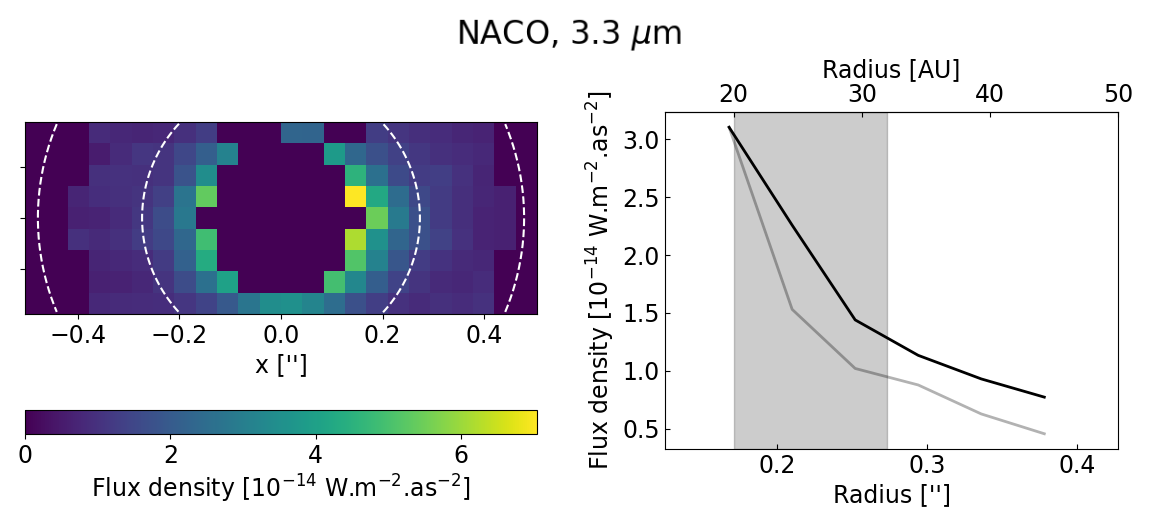}
    \caption{\textit{} Spatial distribution of the emission of HD 169142 observed with NACO, integrated over the 3.3\mic feature (continuum-subtracted, see details  in Sect. \ref{sec:NACOdata}), shown in the upper panel. The inner and outer edge of the gap are shown in with dashed lines. The flux is not computed below 0.168" and above 0.4", due to low S/N. The associated azimuthally averaged emission profile is shown in the lower panel. The black solid line is the radial emission profile at 3.3\mic  and the grey line is the associated PSF. The grey fill indicates the location of micron- and millimeter-sized grain dust that has been taken from the literature.}
    \label{fig:naco}
\end{figure}

The aromatic and aliphatic emission features at 3.3 and 3.4\mic are detected near the star, down to a distance of 0.1", but with a low S/N at such small separations \citep{Bouteraon_032019}. We focus on regions were the features are clearly detected, from 0.168" to 0.4" (20-47~AU) from the star. This corresponds to the transition region between the inner ring and the gap (Fig. \ref{fig:diskstructure}).
For each pixel, the continuum is interpolated by a first-degree polynomial and substracted from the spectrum. To create the pure band emission map and emission profiles at 3.3\mic shown in Fig. \ref{fig:naco}, the spectral region associated to the telluric band between 3.309 and 3.322\mic is removed, and the flux is spectrally integrated between 3.2 and 3.35\mic. 

\section{Analysis of the spatial distribution of the AIB and underlying continuum emission} \label{sec:obsresults}
\subsection{Spatial distribution of AIB carriers}
Figure \ref{fig:VISspectra} shows the VISIR resolved spectra from 0" to 1.02" as well as the HD 169142 spectra integrated over the entire disk, measured by Spitzer, and computed from the VISIR observations. All spectra include both aromatic band emission and dust-continuum emission. The aromatic infrared bands at 8.6 and 11.3\mic are predominant in all VISIR resolved spectra and in the Spitzer spectrum. This confirms that carbonaceous nano-particles dominate the infrared emission in HD 169142, as pointed out by \citet{Seok_2016}. In particular, it shows that the carbonaceous species dominate the infrared emission  in the inner regions from 0" to 1" ($\sim$100~AU). 
The detection of the features at 0", corresponding to a region of a radius of $\sim$0.128" or 15~AU (1 pixel size) centred onto the star, hints towards the presence of carbonaceous nano-particles in the inner cavity. However, we cannot exclude that it is due to the contribution of the inner radius of the ring, located at $\sim$0.17" (20 AU).

A secondary aromatic feature at 12\mic is also detected in the VISIR resolved spectra and Spitzer integrated spectrum. It has already been reported by \citet{Maaskant2013, Seok_2016} in the ISO/SWS and Spitzer/IRS integrated spectra. In the VISIR resolved spectra, this feature is strong in the spectrum centered on the star until a distance of $\sim$0.25" (29~AU) from the star, and decreases at higher radii.

Figure \ref{fig:VISIRphotodata} (left) compares the total emission maps (continuum included) of HD 169142 in the VISIR PAH1 and PAH2 filters to the associated PSF. The full widths at half maximum, computed by a 2D gaussian model fitting, are shown in Table \ref{tab:fwhms} for the PAH1 and PAH2 filters, along with the associated PSF. The emission in the PAH1 and PAH2 filters is spatially more extended than the PSF. Similar observations can be made based on the centro-symmetrically averaged profiles (Fig. \ref{fig:VISIRphotodata}, right, neglecting the disk inclination of $\sim$ \ang{13}). This hints towards an intrinsically extended aromatic emission up to at least 2" (234~AU). The rather flat behaviour of the emission profiles towards the star, and the absence of airy side-lobes in the emission maps and profiles also suggest the need for an inner region (radii below 0.2"), producing relatively low levels of emission in these two bands. A strongly peaked emission towards the inner parts of the disk would indeed lead to noticeable PSF modulations, as described in Sect. \ref{sec:resultsdistribution}.

Figure \ref{fig:naco} shows the spatial distribution of the emission of HD 169142 observed with NACO, integrated over the 3.3\mic feature (continuum-subtracted). The band emission is detected in the inner ring (0.17-0.27", 20-32~AU) and the gap (0.27-0.48", 32-56~AU). \citet{Bouteraon_032019} detected the emission feature at 3.3 and 3.4\mic down to 0.1" with a low S/N. They suggest the presence of a small amount of carbonaceous nano-grains in the inner regions.

\subsection{Estimation of the underlying continuum emission} \label{sec:continuumderivation}
To gain more insight in the intrinsic distribution of the AIB carriers, we aimed at disentangling the pure band emission at 8.6 and 11.3\mic and the underlying continuum emission in the PAH1 and PAH2 filters maps. This requires an accurate estimation of the continuum from the emission in the nearby ArIII and SIV filters, centred at 9 and 10.4\mic. In Fig. \ref{fig:VISspectra}, it can be noted that the continuum emission below the aromatic features at 11.3\mic, and especially at 8.6\mic, shows a spectral slope which vary with distance to the star. Hence, the flux in the ArIII and SIV filters underestimate the continuum emission under each aromatic band. We therefore apply a correcting factor to the flux measured in the ArIII and SIV filters to account for the spectral slopes observed and to derive an accurate continuum flux level at 8.6 and 11.3\mic. The corrected continuum profiles are shown in Fig. \ref{fig:VISIRphotodata}. We derive the correcting factors as follows:
(i) For each spatially resolved spectrum shown in Fig. \ref{fig:VISspectra}, at a radius of $r_i$ = \{0", 0.25", 0.38", 0.64", 0.76", 1.02"\}, we fit the continuum emission by a second order polynomial, and use it to accurately evaluate the continuum flux under each band, $I^s_{cont, 8.6}$ and $I^s_{cont, 11.3}$; (ii) from the same spectra, at the same radii, we measure the flux at the wavelength corresponding to the ArIII and SIV filters, $I^s_{9}$, $I^s_{10.45}$;
(iii) we compute the correcting factors for radius, $r_i,$ as $f_{8.6}(r_i) = \frac{I^s_{cont, 8.6}}{I^s_{9}}$ and $f_{11.3}(r_i) = \frac{I^s_{cont, 11.3}}{I^s_{10.45}}$; (iv) by interpolating the sets \{$f_{8.6}(r_i)$\}$_{i}$ and \{$f_{11.3}(r_i)$\}$_{i}$ over the range of [0", 2"], we derive the correcting factors at all radii $f_{8.6}(r)$, $f_{11.3}(r)$. 

Those factors are then used to compute the underlying continuum maps in the PAH1 and PAH2 filters shown in Fig. \ref{fig:VISIRphotodata}. These maps are estimated by multiplying the ArIII and SIV filters flux by the correcting factors at each radius : $I_{cont, PAH1}(r) = f_{8.6}(r)\,I_{ArIII}(r)$ and $I_{cont, PAH2}(r) = f_{11.3}(r)\,I_{SIV}(r)$, where $I_{X}(r)$ is the flux in the filter X at a radius r.

\subsection{AIB emission versus the continuum}
In this section, we make use of the derivation of the underlying continuum emission in the PAH1 and PAH2 filters (described in Sect. \ref{sec:continuumderivation}) to compare the spatial distributions of the infrared band emission and the continuum emission at 8.6 and 11.3\mic.
As can be seen in Fig. \ref{fig:VISIRphotodata} (right) the total emission (band+continuum) at 8.6 and 11.3\mic and the underlying continuum emission have very similar spatial distributions. The discrepancy between the FWHM of the observational PSF in the PAH1 and PAH2 filters and in the associated continuum filters is smaller than 10\%. Therefore, the comparison between the profiles of the total emission in the PAH1 and PAH2 filters and the underlying continuum can safely be interpreted in terms of the spatial distribution of the emission of the disk. The similarity between the distribution of the total flux and the underlying continuum suggests that the continuum emission at 8.6 and 11.3\mic is dominated by the aromatic infrared band carriers everywhere in the disk. This is in agreement with the dominating aromatic infrared bands in the spatially resolved spectra (Fig. \ref{fig:VISspectra}).

\begin{figure}[!ht]
    \centering
    \includegraphics[width=0.5\textwidth]{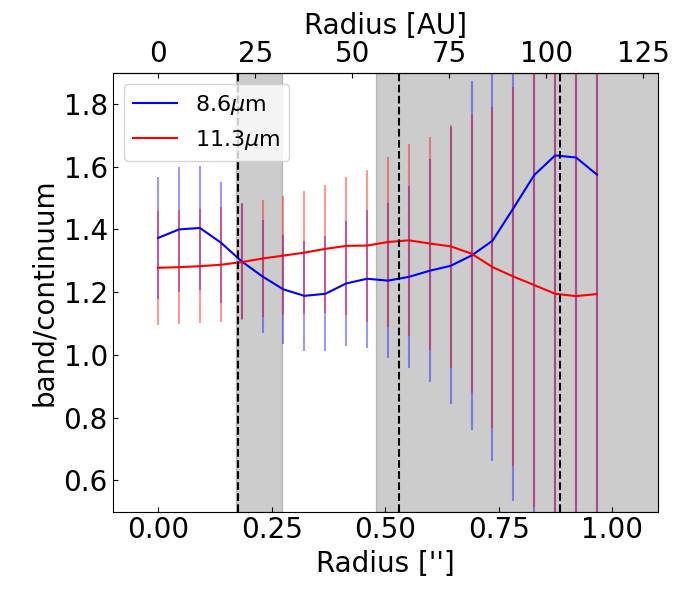}
    \caption{VISIR bands to continuum ratios at 8.6\mic and 11.3\mic. The 8.6 and 11.3\mic band fluxes are azimuthally averaged from the 2D images in the PAH1 and PAH2 VISIR filter (see profiles in Fig. \ref{fig:VISIRphotodata}), and include the continuum and the pure band emission. The derivation of the pure continuum emission is described in Sect. \ref{sec:continuumderivation}. The grey fill indicates the ring and outer disk regions with a continuous distribution of micron- and millimeter-sized grain dust deduced from previous observations (see Fig. \ref{fig:diskstructure}).}
    \label{fig:contratio}
\end{figure}

Figure \ref{fig:contratio} shows the ratios between the total emission and the continuum emission at 8.6 and 11.3\mic. At 11.3\mic, the variations of the ratios are included in the error bars and are not significant. At 8.6\mic, the ratio appears to drop at small radii and increases from 0.75" (0.88~AU), but shows overall little change over 0"-1". These globally flat behaviors of the total emission/continuum ratios are different from what was observed by \citet{Habart2021} in the HD 100546 disk. For HD 100546, the emission near the star is dominated by micronic grains, resulting in an important increase of the band-to-continuum ratios with distance to the star between 0" and 1", and constant ratios at larger distance. The low ratio values closer to the star are mostly due to the increasing thermal emission of the micronic grains at thermal equilibrium (contributing as continuum emission).
On the contrary, the flat band-to-continuum ratios in the HD 169142 disk indicate that carbonaceous nano-grains strongly contribute to the infrared spectra at all radii.

Overall, our data provides evidence of the domination of carbonaceous nano-grains in the infrared emission of the HD 169142 disk at all radii. This is to be linked with the presence of an inner cavity in the disk below 0.17" (20~AU), resulting in a lack of hot micronic grains, which were observed to dominate the continuum emission in, for instance, HD 100546 \citet{Habart2021}.

\subsection{AIB carriers properties} \label{sec:obs_properties}
\citet{Maaskant2014} interpreted the HD 169142 spectrum in the mid-infrared (5-13\mic) using an interstellar-PAH model to account for the aromatic infrared bands and showed that the spectrum is intermediate between fully neutral, for which the 11.3\mic feature dominates over the 6.2, 8, and 8.6\mic features as well as the fully ionized interstellar PAH, showing predominant 6.2, 8, and 8.6\mic features over the 11.3\mic band \citep{Peeters2002}. Following their interpretation, we note that the VISIR spectra show intermediate features between fully ionized and fully neutral at all radii (Fig. \ref{fig:VISspectra}). Nevertheless, as the radius increases, the slope at 8.6\mic flattens, which might suggest a higher neutral fraction in the outer regions of the disk. To highlight this effect, we spatially integrate the spectra within 0.3" (35~AU) and outside of 0.3" separately (see Fig. \ref{fig:VISspectra}, upper panel). The spectrum of the inner regions shows a stronger slope at 8.6\mic than that of the outer regions, hinting towards a higher ionization fraction in the inner parts of the disk. 

According to \citet{Maaskant2014}, the spectrum corresponding to a fully optically thin cavity should show completely ionised features. Nonetheless, it is not the case for the spectrum at 0" (Fig. \ref{fig:VISspectra}), which probes the inner cavity up to 0.128" (15~AU).  The observed neutral carriers emission could originate from the optically thick inner rim of the ring, and be mixed with the emission within 0.128" (15~AU) due to the PSF of the observation. However, it should be noted that the evolution of the spectrum may reflect other effects linked to the irradiation condition (spectral shape and intensity of the UV field) or variations in the properties of the aromatic band carriers, such as their size or their structure (see Sect. \ref{sec:aCproperties_RT}).

\section{Radiative transfer modelling} \label{sec:RTmodels}
 In the following, we use the radiative transfer code POLARIS \citep{Reissl2016}, coupled to the grain model THEMIS \citep{Jones2017} to derive the emission of the disk from a given distribution of small dust grains in HD 169142. Our objectives are to draw better constraints on the distribution of carbonaceous nano-particles in the disk, and to confirm the first preliminary trends provided by the a priori analysis in Sect. \ref{sec:observations}. We mainly focus on the comparison between the radial distribution of carbonaceous nano-grains and that of bigger grains derived from the literature. In particular, we try to constrain the presence of carbonaceous nano-particles in the inner cavity and towards the outer disk.

\subsection{THEMIS dust grain model} \label{sec:THEMIS}
The grain optical properties are modelled using THEMIS, which is an evolutionary core-mantle dust model designed to allow variations in the dust structure, composition, and size according to the local density and radiation field \citep{Jones2013, Jones2017}. The a-C(:H) nano-particles, are modelled as hydrogenated aromatic-rich carbonaceous nano-particles with mixed aromatic, aliphatic, and olefinic domains. It is applied, in particular, to the modelling of dust in the diffuse and dense ISM \citep{Ysard2015, Ysard2016, Saajasto2021}, near-infrared to sub-millimeter observations of nearby galaxies \citep{Chastenet2017, Viaene2019}, the near-infrared observations of protoplanetary disks \citep{Bouteraon_032019, Habart2021}, and in photon-dominated regions \citep{Schirmer2020, Schirmer2021}. 

For the study of the HD 169142 disk, we assume the dust composition and size distribution used by \citet{Habart2021} in their modelling of the HD 100546 disk. It corresponds to the THEMIS diffuse-ISM dust model developed by \citet{Jones2013, Kohler2014}. The dust is modelled as a-C nano-particles populations with different size distributions (a-C$_1$, a-C$_2$, VSG), amorphous carbon dust grains of radius $\sim$ 20~nm (amC), and silicates, modelled as a balanced mix of amorphous pyroxene-type and olivine-type silicate grains (Sil1, Sil2) (see \citet{Habart2021} for more details). The properties of each dust population are given in Table \ref{tab:dustmodel}. The \citet{Habart2021} model corresponds to a weighted mix of the a-C$_1$, a-C$_2$, VSG, and Sil1. It is used in the outer regions of the HD 169142 disk.

The Sil2 dust component is used to model the emission of the HD 169142 inner disk. Similarly to \citet{Chen2018}, we use large grains in the inner structure, with sizes from $10^{-2}$ to $10^3$\mic, and a power-law distribution with exponent -3.5 (see Table \ref{tab:dustmodel}). The higher amount of large grains has already been suggested in inner disk regions, for instance, in the HD 100546 disk by \citet{Tatulli2011}, who used 5\mic -sized dust grains, as well as \citet{Miley2019}, who suggested that grain populations could reach millimeter to centimeter sizes in the inner hot regions of the disk. It is thought to be due to the sublimation of smaller grains in such intense radiation field and temperature conditions \citep{Tatulli2011}. This assumption is physically supported by the fact that the smallest grains are also the hottest ones and are thus more prone to destruction by sublimation.

\begin{table*}[!ht]
    \centering
    \caption{Dust grain populations. The optical properties of the dust populations are computed using the CM20, aPyM5 and aOlM5 components provided by POLARIS as part of the THEMIS model \citep{Jones2013, Jones2017}.}
    \label{tab:dustmodel}
    \begin{tabular}{cccccc}
        \hline
         Name & Composition & Density [kg.m$^{-3}$] & $a_{min}-a_{max}$ [$\mu$m] & Distribution law & (Power-law) exponent/ \\
         &&&&&(Log-normal) mean [$\mu$m]\\
         \hline
         a-C$_1$ & CM20 & 1600 & 4\,10$^{-4}$ - 7\,10$^{-4}$ & power-law & -3.5 \\
         a-C$_2$ & CM20 & 1600 & 7\,10$^{-4}$ - 1.5\,10$^{-3}$ & power-law & -3.5 \\
         VSG & CM20 & 1600 & 1.5\,10$^{-2}$ - 2\,10$^{-2}$ & power-law & -3.5 \\
         amC & CM20 & 1570 &  5\,10$^{-4}$ - 4.9 & log-normal & 7\,10$^{-3}$\\
         Sil1 & $\frac{1}{2}$ aPyM5 + $\frac{1}{2}$ aOlM5 & 2190 & 1\,10$^{-3}$ - 4.9\,10 & log-normal & 8\,10$^{-3}$\\
         Sil2 & $\frac{1}{2}$ aPyM5 + $\frac{1}{2}$ aOlM5 & 2190 &  1\,10$^{-2}$ - 1\,10$^{3}$ & power-law & -3.5\\
         \hline
    \end{tabular}
\end{table*}

\begin{table}[!ht]
\centering
\caption{Total mass in $M_{\astrosun}$ of each dust population depending on the region in the disk.}
\label{tab:dustdistribution}
\begin{tabular}{cccc}
        \hline
         &Inner disk&Inner&Ring+Gap\\
         &&cavity&+Outer disk\\
         \hline 
         Limits [AU]& 0.2-0.5&$R_{cav}^{a-C}$-20&20-200 \\
         \hline
         a-C$_1$&0&0.59\,$M_{cav}^{a-C}$&2.3\,10$^{-6}$\\
         a-C$_2$&0&0.41\,$M_{cav}^{a-C}$&1.61\,10$^{-6}$\\
         VSG&0&0&1.38\,10$^{-6}$\\
         amC&0&0&1.84\,10$^{-6}$\\
         Sil1&0&0&1.59\,10$^{-5}$\\
         Sil2&5.13\,10$^{-13}$&0&0\\
         \hline 
\end{tabular}
\end{table}

\subsection{POLARIS radiative transfer code} \label{sec:POLequations}
POLARIS is a three-dimensional radiative transfer code \citep{Reissl2016} which solves the radiative transfer problem self-consistently using the Monte Carlo method. POLARIS has been developed to simulate polarised emission and Zeeman splitting due to magnetic fields in dusty media. We used the recent updates of the code, which allow for the modelling of the emission of protoplanetary disks by considering complex structures and varying dust populations in space, and take into account the stochastic heating of nanometer-sized grains, including single as well as multi-photon absorption. In particular, feeding POLARIS with optical properties of small carbon grains from THEMIS allows us to build a detailed model of the infrared emission in the HD 169142 disk.

The stellar parameters, R = 1.6\,R$_{\astrosun}$ and T = 7\,800 K, are taken from previous works (Table \ref{tab:HD 169142_star}). We use an axis-symmetrical, cylindrical grid. The gas density distribution is computed assuming hydrostatic equilibrium in the disk \citep{Hayashi1981}:
\begin{equation} \label{eq:evasement}
n(r,z)=n_0 \left( \frac{r_0}{r} \right) ^\alpha exp \left( -\frac{1}{2} \left( \frac{z}{h(r)}\right) ^2 \right)
\text{, } h(r)=h_0 \left( \frac{r}{r_0} \right) ^\beta \text{, }
\end{equation}

where $\alpha$ = - 2.09, yielding decrease in the density with radius slightly steeper than squared, and $\beta$ = 1.1, corresponding to a flared disk, are taken from \citet{Monnier_2017}. We make the commonly used assumption of an approximate $\frac{1}{10}$ ratio for the disk scale height, we set $r_0$= 1\,AU, $h_0$ = 0.07\,AU in the inner disk and we set $r_0$ = 100\,AU, $h_0$ = 15\,AU in the outer regions. The dust model does not include grains bigger than 5\mic, as large grains do not contribute substantially to the thermal emission below 13\mic. The small inclination of the disk is neglected, and the system is modelled face-on, as observed by \citet{Panic2008}, for instance.

Models featuring an empty gap (0.27-0.48", 32-56~AU) or no gap (full disk from 0.17", 20~AU) resulted in similar profiles after convolving them with the VISIR PSF, showing that the VISIR data does not have a high enough angular resolution to study the gap properties (see PSF in Table \ref{tab:fwhms}). Because of the very narrow gap region probed by the NACO data, neither did it allow to distinguish between both cases. For simplicity the gap from 30 to 55 AU is therefore not taken into account in our simulations and the outer region from 20 to 250 AU is modelled as a full disk. Both \citet{Osorio_2014} and \citet{Monnier_2017} place the inner disk between 0.2\,AU and 0.5\,AU, and we follow this choice. The mass of the inner disk is chosen small enough to create an optically thin structure which does not contribute significantly to the disk emission (see Table \ref{tab:dustdistribution}), in agreement with the absence of silicate bump in the HD 169142 spectrum.

The spatial distribution of each type of dust is given in Table \ref{tab:dustdistribution}. We modeled the dust in the outer regions of the disk from 0.17" (20~AU), according to the dust model used by \citet{Habart2021}. It only features dust grains smaller than 5\mic as a mix of a-C$_1$, a-C$_2$, VSG, and Sil1 dust populations (Table \ref{tab:dustdistribution}, third column). The inner disk, below 0.5~AU is modelled as pure silicate dust, with grain sizes up to 1~mm (Table \ref{tab:dustdistribution}, first column). The properties of the dust populations are detailed and discussed in Sect. \ref{sec:THEMIS}. Finally, the model allows the addition of carbonaceous nano-particles below 0.17" (20\,AU) starting from an inner radius $R_{cav}^{a-C}$ and following the gas density distribution given in Equation \ref{eq:evasement} with values for $\alpha$, $\beta$, $r_0$, and $h_0$ identical to the outer disk. This new component is composed only of carbonaceous nano-particles (a-C$_1$, a-C$_2$, Table \ref{tab:dustdistribution}, second column) with a density depleted by a factor $f$ compared to the outer disk, resulting in a total carbon nano-particles mass $M_{cav}^{a-C}$ between $R_{cav}^{a-C}$ and 20\,AU. The numerical values used for the model parameters ($h_0$, $R_{cav}^{a-C}$, $M_{cav}^{a-C}$) are given in Table \ref{tab:Mimodels} and will be discussed in Sect. \ref{sec:resultsdistribution}.

\begin{table}[!ht]
\centering
\caption{Properties of the POLARIS models for the HD 169142 disk. $h_0$ is the disk height à 100\,AU, $R_{cav}^{a-C}$ is the inner radius of the aromatic particles in the cavity, $M_{cav}^{a-C}$ is the total mass of aromatic particles below 0.17" (20\,AU).}
\label{tab:Mimodels}
\begin{tabular}{cccc}
        \hline
         Model&$h_0$&$R_{cav}^{a-C}$&$M_{cav}^{a-C}$\\
         &[AU]&[AU]&[$M_{\astrosun}$]\\
         \hline 
         M0&15&20&0\\
         M1&5&20&0\\
         M2&5&5&3.56\,10$^{-4}$\\
         &&&(f=1)\\
         M3&5&5&1.19\,10$^{-7}$\\
         &&&(f=3000)\\
         \hline
\end{tabular}
\end{table}

\subsection{Predicted intensity profiles}
POLARIS outputs are obtained as data cubes: 2D intensity maps computed at regularly spaced wavelengths (150 values between 7 and 13\mic). To account for the spectral width of VISIR PAH1 and PAH2 filters in our simulations, the maps are spectrally averaged over the filter widths. The resulting images are then convolved with the observed PSF (given in Fig. \ref{fig:VISIRphotodata}) and azimuthally averaged to compute the profiles shown in Fig. \ref{fig:POLVISIR}. The continuum maps and profiles are computed first by a two-degree polynomial interpolation of the continuum around 8.6 and 11.3\mic on the initial POLARIS cube, resulting in a continuum cube, from which the continuum profiles are then similarly derived.

Similar work has been done to reproduce the NACO maps and profiles to account for the spectral width of the aromatic band at 3.3\mic. The simulated spectral cube (100 wavelengths between 3 and 4\mic) is first spectrally integrated between 3.2 and 3.35\mic, ignoring the region between 3.309 and 3.322\mic, which corresponds to a telluric band. Ignoring the telluric band results in a underestimation of the modelled integrated band flux by less than 0.5\%. We then extract a 1D flux profile and convolve it with the PSF (1D, see Fig. \ref{fig:naco}) of the observation, integrated in the same way around 3.3\mic. The resulting profiles are compared with NACO data in Fig. \ref{fig:POLNACO}.

\section{Comparison between model and data} \label{sec:results}
\subsection{Constraining the nano-grain distribution in the disk} \label{sec:resultsdistribution}
\begin{figure*}[!ht]
    \centering
\includegraphics[width=0.8\textwidth]{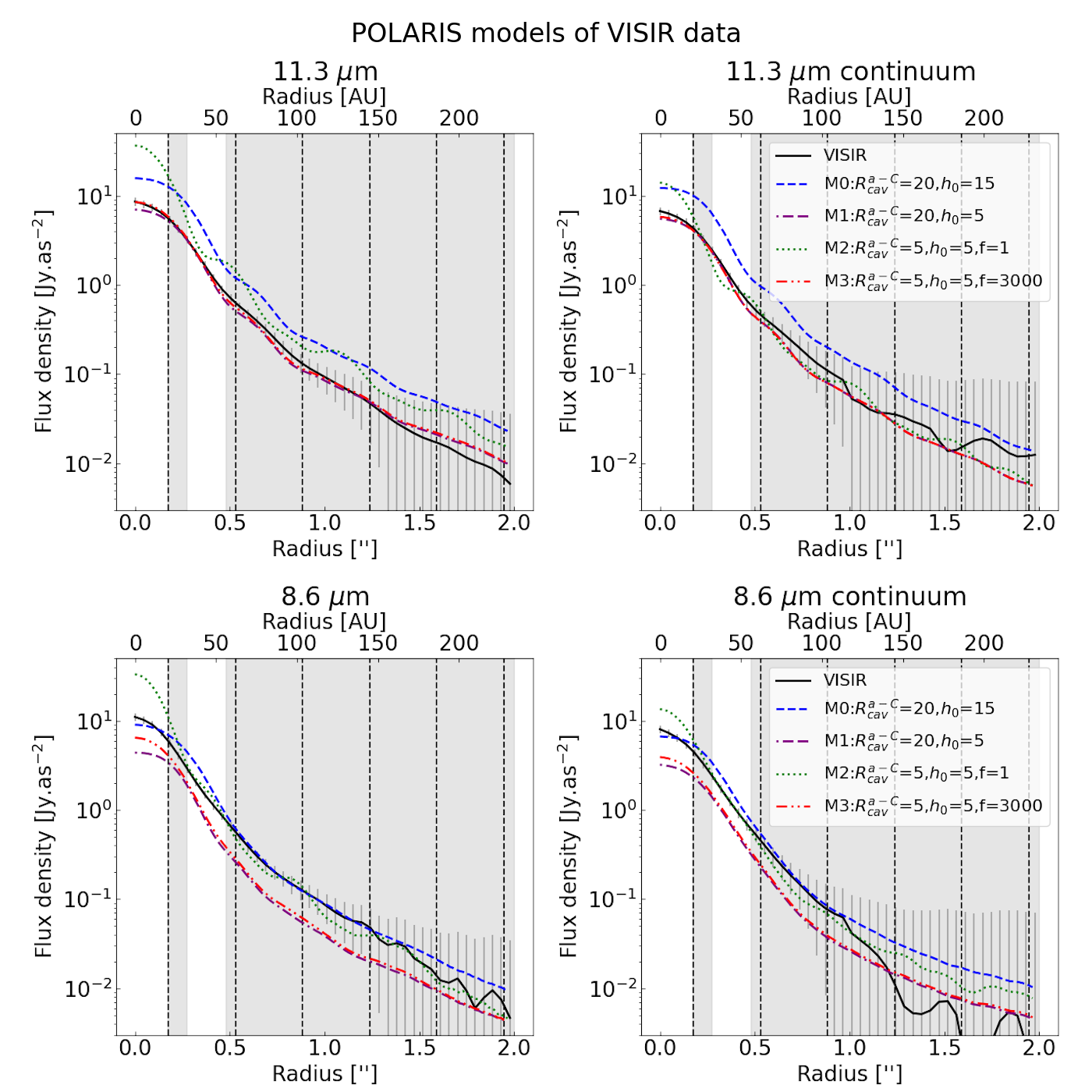}
    \caption{VISIR radial profiles (black full line), compared to simulated results (dotted, dashed, dash-dash-dotted, and dash-dot-dotted lines) for the PAH2 filter and associated continuum on the top panels, and the PAH1 filter and continuum on the bottom panels. For a description of the different models, we refer to Table \ref{tab:Mimodels}. The grey fill indicates the location of micron- and millimeter-sized grain dust.}
    \label{fig:POLVISIR}
\end{figure*}

We first compared the data with the standard disk model described in Sect. \ref{sec:POLequations}, having no carbonaceous nano-particles below 0.17" (20\,AU), as shown in model M0 in Table \ref{tab:Mimodels} and Fig. \ref{fig:POLVISIR}. Figure  \ref{fig:POLVISIR} (top) compares the resulting profiles with the VISIR imaging data in the 11.3\mic PAH2 filter. We focus mainly on the model-data comparison in this filter, because it is less affected by the state of charge of the carriers. The VISIR flux is overestimated at 11.3\mic in the entire disk by a factor of 1.8 to 2.2, both for the continuum and the total emission, resulting in an overestimation of the band flux by a factor of 1.9 over the full disk (Table \ref{tab:integrfluxes}).

This overestimation at all radii can be linked to the presence of a spatially unresolved component showing very strong emission from all dust populations in the inner regions in our model. Such a component would indeed contribute as an additional, almost unresolved flux at every radius in the disk.
We lower the value of the disk height in the whole disk by decreasing $h_0$ from 15 to 5~AU at $r_0$ = 100~AU. This allows to reduce the amount of hot dust close to the star and recover the VISIR total flux level in the PAH2 filter also as the associated continuum (model M1 in Table \ref{tab:Mimodels} and Fig. \ref{fig:POLVISIR}) and the integrated continuum-subtracted emission in the band (Table \ref{tab:integrfluxes}). 
The slope of the modelled profile agrees with the VISIR data in the PAH2 band for radii higher than 0.2" (Fig. \ref{fig:POLVISIR}) and the absolute flux at 11.3\mic is well reproduced. However, the modelled flux in the 11.3\mic feature drops by 20\% (1.8~Jy.$as^{-2}$) below the VISIR error bars at radii smaller than 0.2" (23~AU). 

Adding a very small amount of carbonaceous nano-particles in the inner cavity leads our model to match the entire observed profile within the error bars (model M3 in Table \ref{tab:Mimodels} and Fig. \ref{fig:POLVISIR}). However, a degeneracy is observed between the inner radius, $R_{cav}^{a-C}$, and the depletion factor, $f,$ of the carbonaceous nano-grains in the cavity. This is because models using ($R_{cav}^{a-C} = 0.128 \text{" or } 15~\text{AU}; f=100$), ($R_{cav}^{a-C} = 0.09 \text{" or } 10~\text{AU}; f=1000$) and ($R_{cav}^{a-C} = 0.04 \text{" or } 5~\text{AU}; f=3000$) account similarly well for the VISIR profiles at 11.3\mic, the ISO, and the Spitzer spectra. A model having non-depleted carbonaceous nano-grains in the cavity from 0.04 to 0.17" (5~AU to 20~AU) strongly overestimates the flux of the aromatic infrared bands (model M2 in Table \ref{tab:integrfluxes}). The associated profile at 11.3\mic overestimates the band flux near the star, and shows Airy side-lobes (Fig. \ref{fig:POLVISIR}). The much better agreement achieved between model M3 and the data thus supports the need for depletion of carbonaceous nano-particles in the cavity.

In summary, the comparison between our models and the VISIR data at 11.3\mic hints towards depleted carbonaceous nano-grains in the cavity below 0.17'' (20~AU). It could be due to photo-destruction of the grains in the strongly irradiated inner regions. The photo-destruction and depletion of small dust grains has been observed, for instance, in the interstellar photon dominated regions by \citet{Schirmer2020}. However, this study faces strong degeneracy due to the limited angular resolution of the data. A higher angular-resolution is required to better quantify the level of depletion as well as the spatial distribution of the band carriers in the cavity.

\subsection{Carbonaceous nano-particles properties} \label{sec:aCproperties_RT}
\subsubsection{Analysis of the 8.6\mic band}
\begin{figure}[!ht]
    \centering
    \includegraphics[width=0.5\textwidth]{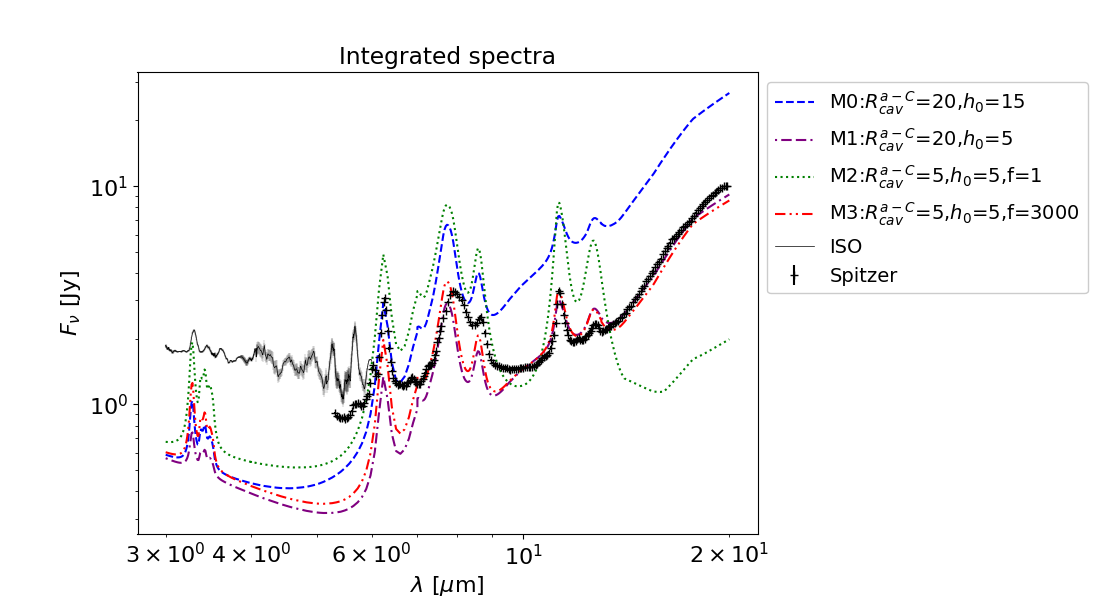}
    \caption{ISO and Spitzer spectra, compared to the simulated results. For a description of the different models, we refer to Table \ref{tab:Mimodels}.}
    \label{fig:POLspectr}
\end{figure}

None of the models  have satisfactorily reproduced the VISIR 8.6\mic band profile (Fig. \ref{fig:POLVISIR}). Both the M1 and M3 models, which optimally reproduce the data at 11.3\mic, underestimate the flux at 8.6\mic by a factor 2 in the entire disk, as well as in the integrated spectra (Fig. \ref{fig:POLspectr}). 

This could be explained considering the charge state of the band carriers. Using an interstellar-PAH description of the aromatic infrared band carriers, the 8.6\mic feature is expected to be stronger for ionised PAH. The underestimation of the emission at 8.6\mic would then suggest the need for a higher ionisation fraction in the HD 169142 disk than what is considered in our model. Indeed, THEMIS includes no charge dependence of the aromatic particles, while \citet{Seok_2016} derived an ionization fraction of 0.6 for the PAH in HD 169142 from SED fitting. 

In addition, all simulated VISIR profiles at 8.6\mic are too flat near the star and show a strong slope between 0.2" and 0.5" (20-60~AU). In order to better fit the data, the predicted emission of the 8.6\mic feature should be stronger below 0.17" (20~AU) and quickly decrease when entering the disk. It is in agreement with the evolution of the 8.6\mic slope observed in the VISIR resolved spectra (see Fig. \ref{fig:VISspectra} and Sect. \ref{sec:obs_properties}). In the interstellar-PAH framework, as the 8.6\mic to 11.3\mic ratio is higher for ionised PAH, it thus suggests the need for a greater amount of ionised emission near the star in our disk models. As the POLARIS/THEMIS models assume no variations of the ionisation fraction, the comparison to the data would then hint towards an increasing ionisation fraction towards the irradiated inner regions. \citet{Maaskant2014} associated the emission from ionized PAH to optically thin inner disk regions where the UV field is high and the electron density is low; whereas optically thick outer disk regions would be dominated by neutral PAH emission with little variation with radius. This is in agreement with the trend found in our observations. 

However, the interpretation of the spectral signatures at 7.7, 8.6, and 11.3\mic is not unequivocal, and could be as well explained, for instance, by varying size distribution of carbonaceous nano-grains. The underestimation of the flux at 8.6\mic, and the need for a steeper radial slope near the star would, within this framework, suggest the need for a higher abundance of bigger carbonaceous dust grains in the inner regions. This is consistent with stronger photo-destruction of smaller grains in optically thin, strongly irradiated media and is supported by the work of \citet{Schirmer2020, Schirmer2021} on dust evolution in photon dominated regions.

\subsubsection{Analysis of the 3.3\mic band profile}
\begin{figure}[ht]
    \centering
    \includegraphics[width=0.4\textwidth]{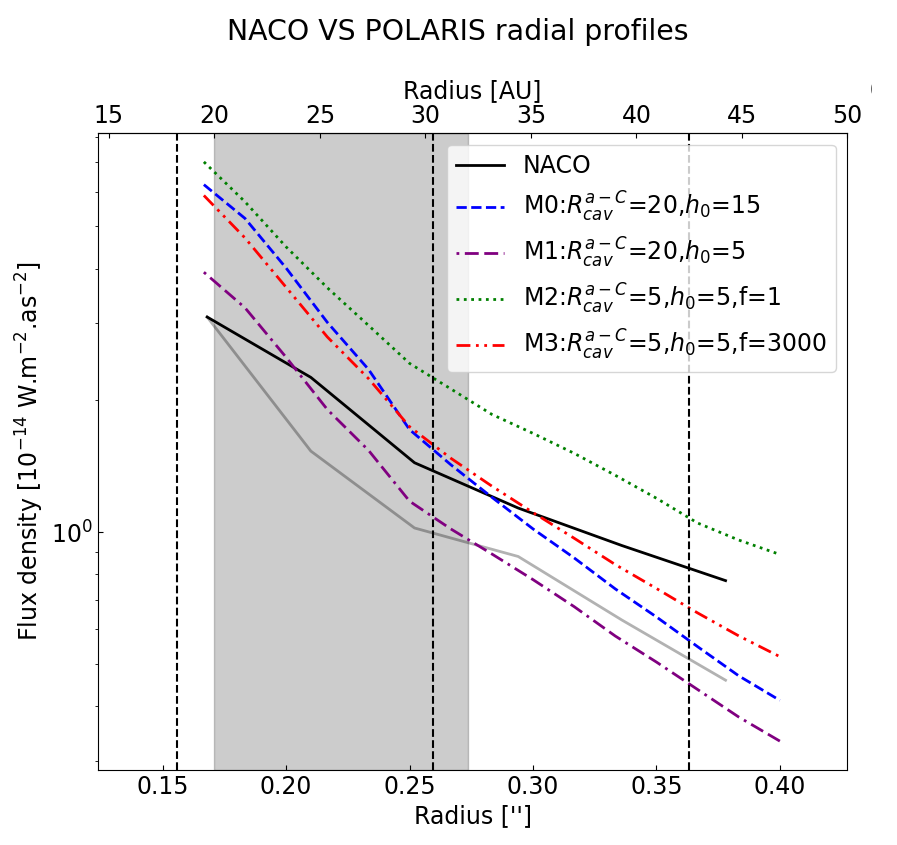}
    \caption{NACO emission profiles (black line) and PSF (grey line) compared to POLARIS simulated results (dotted, dashed, dash-dash-dotted and dash-dot-dotted lines). For a description of the different models, see Table \ref{tab:Mimodels}. The grey fill indicates the location of micron- and millimeter-sized grain dust.}
    \label{fig:POLNACO}
\end{figure}

The standard model M0 as well as models M1 and M3 are consistent with the NACO absolute flux in the 3.3\mic band (Table \ref{tab:integrfluxes}). Yet, none of the simulated flux distributions agree with the NACO radial profile shown in Fig. \ref{fig:POLNACO}. The 3.3\mic is produced by very small carbon grains, with radii smaller than 0.7~nm. A stronger depletion of this population of grains in the innermost regions could explain the discrepancy between the data and the model. Such a depletion could originate from photo-destruction of the smaller grains in the irradiated regions, as was observed in the interstellar photon dominated regions by \citet{Schirmer2020}.

\begin{table*}[ht]
    \centering
    \caption{Continuum-subtracted band fluxes in $10^{-14}\,W.m^{-2}$, integrated over the disk in the three aromatic bands, computed for the different the POLARIS models. The modelled values are compared to the fluxes computed on ISO data \citep{Acke2004} and Spitzer data \citep{Acke2010}.}
    \label{tab:integrfluxes}
    \begin{tabular}{cccc}
        \hline
        Band &3.3\mic&8.6\mic &11.3\mic \\
        \hline
        \multicolumn{4}{c}{\underline{Data}}\\
        Spitzer&&0.89 $\pm$ 0.01&2.08 $\pm$ 0.01\\
        ISO&1 $\pm$ 0.2&2.55 $\pm$ 0.5&2.3 $\pm$ 0.5\\
        \multicolumn{4}{c}{\underline{Model}}\\
        \textbf{M0} : $R_{cav}^{a-C}$=20, $h_0$=15&0.55&1.3&2.2\\
        \textbf{M1} : $R_{cav}^{a-C}$=20, $h_0$=5&0.26&0.64&1.1\\
        \textbf{M2} : $R_{cav}^{a-C}$=5, $h_0$=5, f=1&1.7&3.5&7.8\\
        \textbf{M3} : $R_{cav}^{a-C}$=5, $h_0$=5, f=3000&0.87&0.92&1.5\\
        \hline
    \end{tabular}
\end{table*}

\section{Conclusion} \label{sec:conclusion}
We investigated the carbonaceous nano-grains emission in the HD 169142 disk by comparing high angular observations with model predictions. We analysed the imaging and spectroscopic
 data (VLT/VISIR, 8-12\mic, angular resolution 0.3") as well as the adaptive optics spectroscopic data (VLT/NAOS-CONICA, 
3-4\mic, angular resolution 0.1"). Our main results derived from both observations and model predictions are summarised below.
 
\begin{enumerate}
    
\item Probing the emission in the HD 169142 disk in the 3.3, 8.6, and 11.3\mic aromatic infrared bands, we were able to provide new strong evidence of the extended nature of the distribution of carbonaceous nano-particles populations and their domination in the infrared emission of the disk at all radii, in agreement with previous studies, for instance \citet{Seok_2016}. 
    
\item The flattening VISIR spatial emission profiles near the star, the absence of Airy side-lobes in the VISIR imaging data and the faint aromatic feature detection by NACO below 20~AU points towards a fairly large depletion of PAH inside the 20~AU cavity. 

\item To gain more insight in the spatial distribution of carbonaceous nano-particles in the disk, we modelled the distribution of small dust grains using the POLARIS radiative transfer code coupled to the  THEMIS model for the emission of carbonaceous nano-grains.
The VISIR radial profile at 11.3\mic  is reproduced within the error bars, assuming a continuous radial distribution of micronic grains and carbonaceous nano-grains from 20 to 250\,AU, as well as carbonaceous nano-grains decreased in abundance in the inner cavity between 0.5 and 20\,AU. The precise amount of carbon nano-grains needed in the cavity and their inner radius are strongly degenerate and could not be derived from our modelling. Nonetheless, the comparison between simulations and the emission profile at 11.3\mic shows the need for strongly depleted carbonaceous nano-grains below 20\,AU. This result supports previous studies from \cite{Maaskant2013} and \citet{Seok_2016}, which hinted at the presence of interstellar PAH in the cavity and offers evidence of their depletion.

\item The VISIR spectra clearly shows a spectral evolution from the inner regions towards the outer regions. In the interpretation framework of interstellar PAH, this evolution can be explained by a transition between dominating ionized PAH in the optically thin internal regions to neutral PAH in the optically thick outer disk media.
The discrepancy between our models and the VISIR profiles at 8.6\mic seems to support this interpretation. It first suggests that the ionization which has be considered in the model is too small (full neutrality was assumed). As the emission at 8.6\mic shows more radial variations in the data than in the model, it also suggests the need for ionization variations throughout the disk. However, the interpretation of the spectral signatures at 8.6 and 11.3\mic as indicators of the ionization fraction of the PAH is not unequivocal. The evolution of the spectrum may reflect other effects linked to the irradiation condition (spectral shape and intensity of the UV field) or the size or structure of the band carriers. The evolution of the HD 169142 spectrum towards the center could also be consistent with a higher abundance of larger grains in the inner cavity. 

\item Finally, the NACO radial slope is strongly overestimated. This could be linked to a depletion in very small carbonaceous nano-particles towards the most irradiated regions, caused by a stronger photo-dissociation of small particles \citep{Schirmer2020}. 
\end{enumerate}

Our understanding of the evolution of carbonaceous nano-grains in disks is expected to make significant progress thanks to upcoming spatially resolved observations with the VLTI/MATISSE interferometer, the JWST integral-spectrograph, and the ELT/METIS high-resolution spectrograph and high-contrast imager.
Interferometric measurements using MATISSE would provide valuable and precise information on the distribution of carbon nano-grains in the innermost regions, namely, the inner cavity and the inner disk (1 $<$ r $<$ 15~AU). 
On the other hand, observations with the JWST covering a complete spectral domain between 0.6 and 28~$\mu$m will mainly probe the dust content with a spectral resolution of $\sim$3000-4000, a spatial resolution of 10-100~AU, and for distances up to 500~AU. Spatially resolved spectroscopy of the aromatic and aliphatics sub-features will be obtained and allow for a better identification of the nature and properties of the bands carriers. 
Finally, ELT-METIS will achieve a combination of high spatial resolution with a large field-of-view and a spectral resolution allowing the detection of aromatic infrared bands, thus providing an great opportunity to probe the dust content in the gap region (0.27"-0.48" or 32-56~AU).
Furthermore, combining dust models with new high-resolution data could allow a more detailed analysis of the local properties of the particles (charge, size, excitation). This would help to break some of the degeneracies encountered in this work, \ differentiate between the various proposed disk compositions, and study more precisely the distribution of aromatic dust populations in the gaps.

\begin{acknowledgements} 
We thank the anonymous referee for enlightening comments. This work was based on observations collected at the European Southern Observatory, Chile (ESO proposal number: 075.C-0624(A)), and was supported by P2IO LabEx (ANR-10-LABX-0038) in the framework of the “Investissements d’Avenir” (ANR-11-IDEX-0003-01) managed by the Agence Nationale de la Recherche (ANR, France), P2IO LabEx (A-JWST-01-02-01+LABEXP2IOPDO), and Programme National “Physique et Chimie du Milieu Interstellaire” (PCMI) of CNRS/INSU with INC/INP co-funded by CEA and CNES.
\end{acknowledgements}

\bibliographystyle{aa}
\bibliography{biblio.bib}

\end{document}